# Virtual Laboratories in Cloud Infrastructure of Educational Institutions

Evgeniy Pluzhnik
Moscow Technological Institute
Moscow, Russia, Leninckiy pr,, 39A, 119334,
e.pluzhnik@gmail.com

Evgeny Nikulchev
Moscow Technological Institute
Moscow, Russia, 119334, Leninckiy pr., 39A
nikulchev@mail.ru

*Abstract*— **Modern educational institutions widely used virtual laboratories and cloud technologies. In practice must deal with security, processing speed and other tasks. The paper describes the experience of the construction of an experimental stand cloud computing and network management. Models and control principles set forth herein.**

## I. Introduction

The report focuses on research in the field develop the principles of database design in cloud hybrid systems for high-quality service of BigData [1].

Currently solved the following local tasks:
- Design a Laboratory Test Bench
- Model a test database
- Develop real time control principles
- Conduct experiments
- Introduce criteria for optimal control and effective activity
- Define problems which provide a solution of the global task

To form a block of workflow management system to query scientific and educational content of the experimental studies, this can be schematically represented as in Fig. 1.

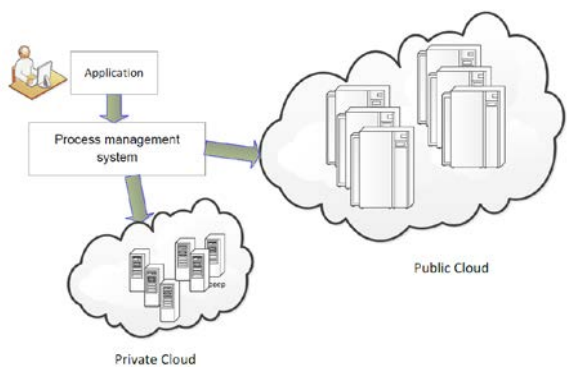

Fig. 1. Management system hybrid cloud systems

To build information systems research and educational content with semistructured data necessary to solve the following problems:
- Evaluation of the general parameters of the system (maximum number of users for simultaneous operation, the possibility of scaling services, the availability of personalized access).
- Project cost estimate (having our own server capacity, cost comparison with the cost of building rental services).
- Evaluation time data access, query performance evaluation for cloud infrastructures.
- Construction of automatic allocation system and send requests in a distributed database

Note that when migrating to the cloud infrastructure are experiencing the exact same multiplayer server systems, such as MATLAB [2, 3].

## II. Laboratory Test Bench

Laboratory Test Bench Functionally test bench can be divided into three main groups, allow you to create multi-functional system for each test problem. Boxing Statistics and management consists of: system of gathering and storing; statistical data management system; generation traffic system [4, 5].

Computing architecture given in Fig. 2.

*Server virtualization*. To conduct research on the server SunFire X2200M2 installed VMware ESXi as the place of installation used flash memory is configured virtual switch Cisco Nexus 1000 and deployed 4 virtual machines on a physical server's disks.

*Physical server*. On HP Proliant DL320 installed Microsoft Windows operating systems and collectors collection of system parameters.

*Organization of the cloud*. For the cloud is used product family VMware vCloud, allowing to organize cloud computing at all three levels. To create clouds in the experimental stand on two servers SunFire created hosts VMware ESXi system installed VCenter management, installed VMware vCloud Director, using MSSQL database to manage the cloud, as vApp applications are created virtual server with the installed applications.

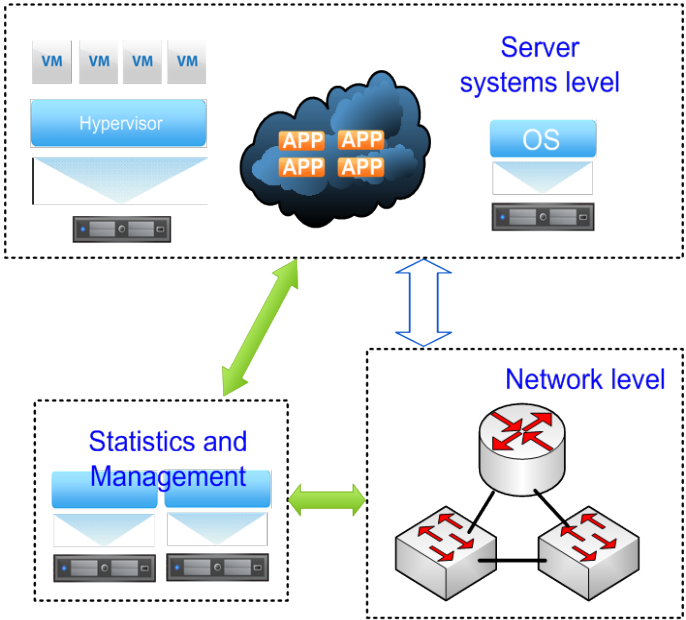

Fig. 2. Laboratory Test Bench

Software is used VMware vCloud, VMWare vSphere, operating systems and database Microsoft, FreeBSD.

The presence of more than 15 physical switches Cisco 29 series routers and 26 and 28 of the series, as well as virtual switches Nexus, functional network equipment, the use of dynamic routing protocols, technology Vlan, trunk, QoS and other allows to implement various schemes network for a broad range of tasks.

Existing functionality allows you to conduct research in the following directions:
- analysis of the traffic routing and switching;
- dynamic control downloads channels of communication, QoS priorities;
- dynamic balancing of traffic and load, a study reservation systems;
- analysis of bottlenecks and problem areas in the network part of hardware and software.

## III. BASIC CONTROL PRINCIPLES

The base model offers classic model is approximated by the following difference equations

$$x(t+1) = Ax(t) + Bu(t),$$

where $x = (x_1, x_2, ..., x_n)^T$ — n-dimensional vector of the system states under given constraints $x \in X \subseteq R^n$, $u = (u_1, u_2, ..., u_m)^T$ — m-dimensional vector of $u \in U \subseteq R^m$.

Basic control principles is:
- Methods of based on single-step Control System Model are used
- Each block of a distributed system in the hybrid cloud infrastructure is a subsystem
- All subsystems exchange signals being control signals, requests, data and references to data
- Feedback from subsystems distributed in the infrastructure is needed

For computing systems are intuitive terminologies of structural complexity, observability, reachability of these information systems [6]. That means all of the many well-developed tools and control theory can be applied to study the parameters of information systems.

Given the logic of the cloud computing is difficult to determine the effectiveness of the algorithms. Virtualization and scalability of resources may unexpectedly greatly speed up the algorithms, as in our experimental example. That's a probably will help identify methods of system identification.

In general, it can be noted that the transfer system is necessary to develop a stable structure, the study of the characteristics of reliability, observability. Decomposition of the local and cloud components should be based on the methods of structural complexity of systems analysis.

The dynamic descriptions in the form of approximation of differential equations with the vector control were used in the different tasks that are close to the considered in the article. In [7] the processes that include real-time embedded systems. You can also note the earlier article for managing web servers [8], and virtualized data centers [9].

In previous works, such as [10], an approach that assumes that the cloud can be modeled as a Multi-Input-Multi-Output (MIMO), the system is implemented for capacity control in the cloud. However, the design of the regulars and identification systems do not allow the full use of these ideas. In the work [11] proposed to use Single-Input-Single-Output (SISO) systems with the introduction of the notion proportional thresholding.

The original paper [12] is devoted to the construction of a dynamic model of linked servers in the cloud. The important results of this work are the introduction of the concept of positive feedback, the theoretical proof of the stability of the model, the use of passive systems.

Dynamic issues are considered and applied to ensure QoS. In [13] discusses the use of a guaranteed rate transmission mechanism for maintaining QoS during network congestion. In [14] the development of a routing protocol to ensure QoS using temporary bandwidth reservation. In [15] deals with the modeling of traffic for the mobile network based on QoS.

## IV. CRITERIA

Optimality criteria:

1. Absence of delays in receipt of data in the course of migration of a Computing Systems to the hybrid cloud.

2. Decrease of the amount of duplicated request under peak load conditions (workload of channels, processors in the cloud.

Estimation of a test system by criterion 1. The interaction time practically did not change during the migration [1]. For BigData – the interaction time in the hybrid structure decreased

(due to increase of resource capacities), which compensated the losses on network data transfer

Estimation of a test system by criterion 1. Before dynamic data flow control the local system part workload histogram had a heavy-tailed distribution (Fig. 3).

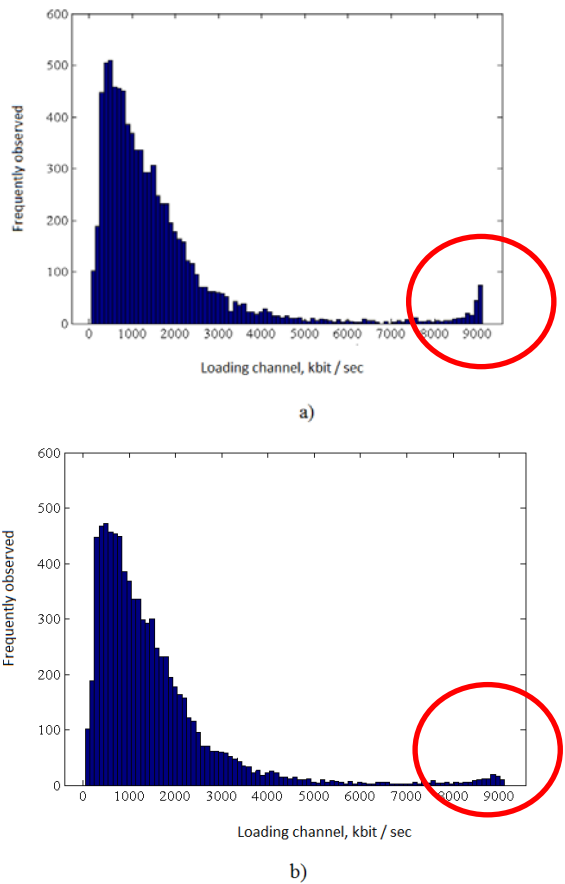

Fig. 3. Distribution: a) no control; b) with control

After the control "tails" in the distribution straightened [5]. This points to the fact that the number of duplicated requests decreased under peak load of resources.

V. DIRECTIONS FOR THE DEVELOPMENT OF STUDIES

Develop formal approaches to ensuring of structural stability of decomposition during migration at the stage of system design.

Develop recommendations on migration of large systems into hybrid clouds and ensuring guaranteed performance thereof.

Develop principles of dynamic data transfer control with due account for feedback on data processing status.